# An Approximation for Random Surfaces on Arbitrary Target Spaces

Mark Wexler

*Laboratoire de Physique Théorique et Hautes Energies*
*Universités Pierre et Marie Curie (Paris VI) et Denis Diderot (Paris VII)*
*Boite 126, 4 place Jussieu, 75252 Paris cedex 05, France*
*wexler@lpthe.jussieu.fr*

**Abstract**

A perturbative technique, the low-temperature expansion, is developed for matrix models of random surfaces. It can be applied to models with arbitrary target spaces, including ones with $c > 1$. As a simple illustration, the series is worked out to 10th order for the surface coupled to a $q$-state Potts model. Accurate estimates for, e.g., $\gamma_{str}$ are obtained both in the low $q$ ($c < 1$) and high $q$ (branched polymer) regimes, including the logarithmic corrections to scaling.

## 1 Introduction

The trouble with matrix models of 2D random surfaces and quantum gravity [1] is that "we are in the unfortunate situation of either solving exactly (for $c \leq 1$) or not understanding at all what is happening" [2]. If the target space has loops, we do not even know where to begin.

In this paper I develop a perturbative technique, the low temperature expansion (LTE), which can handle any multi-matrix model, and therefore any kind of target space or matter model on the surface [3, 4]. In effect, the LTE expresses multi-matrix model correlation functions in terms of one-matrix model correlation functions. The result is a series in powers of $e^{-\beta}$ ($\beta$ is the inverse matter temperature), whose coefficients depend on the cosmological constant, $g$, and on the geometry of the target space (which I will always take to be a graph on a discrete set of points). In fact, most of the calculations required to obtain these coefficients are universal—they are identical for any target space. The target space enters at only one step, in the calculation of so-called "homomorphism coloring factors."

There exists another series technique for matrix models, the strong-coupling expansion, where the expansion parameter is the cosmological constant [5, 6, 7]. There, the $n$-th term includes all surfaces made up of $n$ simplices. The difficulty of the strong-coupling approach—which has been pushed to about 10th or 20th order, depending on the target space—is that one cannot always trust extrapolations from such small surfaces, especially above $c = 1$, where the geometry becomes quite singular.

The LTE does not suffer from this difficulty. In every order it includes surfaces of *all* sizes. The LTE is a spin cluster expansion: what is constrained is not the total area of the surface, but the number of spin clusters.[1] With very moderate computational effort it gives accurate results for the critical points and exponents (including the logarithmic corrections) for $c \leq 1$, as well as for the (even more) branched large-$c$ surfaces.

In the present article, following a detailed discussion of how to calculate the LTE[2], I carry the series out to 10th order for the simplest multi-matrix model, the $q$-state Potts model.[3] The $n$-th term of the LTE, the coefficient of $e^{-n\beta}$, is now an explicit function of $g$ and $q$. In other words, one does *not* have to do a new calculation for each new value of $q$—once the calculation of the LTE is finished, one simply plugs in values of $q$ into the series coefficients. The Potts example is meant foremost as an illustration of the effectiveness of the LTE; but also as a study of a model interesting in its own right. What follows is a brief discussion of the Potts model on a random surface: first of all, why is it interesting?

A standard argument for why it's *not* runs as follows: to get a non-trivial modification of pure gravity, one must take a model which—on a *fixed* 2d lattice—has a continuous transition, and couple it to the surface. But for $q > 4$, the Potts model has only discontinuous transitions on a *fixed* lattice. Doesn't this mean that putting a $q > 4$ model on a random surface should yield only pure gravity? No, because coupling a spin model to a random surface (usually) softens the spin model's phase transition. For instance, the transition of the $q = 2$ Potts model (the Ising model) is second order ($\alpha = 0$) on a fixed 2d lattice, and third order ($\alpha = -1$) on a random surface. The effect is even more dramatic at $q = \infty$: on a fixed lattice the model is *very* first-order, while on a random surface it is *third*-order. At $q = \infty$ the model is in fact exactly and explicitly solvable [8]; surprisingly, it is identical to $c = \infty$ models (such as the $d = \infty$ lattice, or infinitely many Ising spins) [4]. This, then, is the interest of the Potts model on a random surface: for $q \leq 4$ it corresponds to $c \leq 1$, while for $q = \infty$ it is identical to $c = \infty$. Can the $4 < q < \infty$ region teach us anything about $c > 1$? And if there are continuous phase transitions for $q > 4$ that modify pure gravity, do these correspond to some new flat models coupled to gravity, or are these models unique to random surfaces?

---

[1] The number of links between spin clusters, to be exact; but the number of clusters is never greater than one plus the number of links between them.

[2] The LTE was introduced in [3], but as a qualitative rather than quantitative method. In [4, 8] it was resummed to all orders in the $q, c, d \to \infty$ limit.

[3] "Simplest" because its target space, $K_q$, has no structure whatsoever: every point (or matter state) is connected to every other point, with the same weight. Consequently, the homomorphism coloring factors reduce to ordinary chromatic polynomials—see section 2.5.

The model is defined as follows. There are two kinds of degrees of freedom: the lattice and the matter. The lattices are randomly triangulated 2D spherical surfaces, which are dual to spherical cubic fatgraphs. The matter consists of a spin $\sigma_a \in \{1, \ldots, q\}$ on each triangle $a$ (or on each vertex of the fatgraph). For a given geometry, the matter Hamiltonian is

$$\mathcal{H} = \sum_{e_{ab}} e^{\beta(\delta_{\sigma_a \sigma_b} - 1)} \tag{1}$$

where the sum runs over all edges of the fatgraph (a triangle may be connected to itself, and two triangles may be connected by more than one edge). The partition function of the model is realized by the following matrix integral (putting $e^{-2\beta} = z$):

$$F_q(g, z) = \frac{1}{q} \log \frac{1}{f_0} \int \mathcal{D}\phi_1 \cdots \mathcal{D}\phi_q \exp -N \operatorname{Tr} \left( \frac{1}{2} \sum_{ij} \phi_i (T^{-1})_{ij} \phi_j + g \sum_i \phi_i^3 \right) \tag{2}$$

where the $\phi_i$ are $N \times N$ hermitian matrices and the matrix integral is to be understood as being divided by $N^2$ and taken in the limit $N \to \infty$. The matrix $T$ defines the matter that is to be coupled to the fluctuating surface—it is the connection matrix of the target space graph; in the case of the Potts model this is

$$T_{ij} = \begin{cases} 1 & i = j \\ z & i \neq j \end{cases} \qquad (T^{-1})_{ij} = \begin{cases} \frac{1+(q-2)z}{(1-z)(1+(q-1)z)} & i = j \\ \frac{-z}{(1-z)(1+(q-1)z)} & i \neq j \end{cases} \tag{3}$$

It will be convenient to use a rescaled coupling constant $a = \Delta z$, $\Delta = q - 1$. Finally, the normalization $f_0$ is defined by requiring $F_q(0, z) = 1$.

What do we expect to find? First, for every value of $q$ there should be a curve of critical points in the $(g, a)$ plane. For $q \leq 4$ ($c \leq 1$) we know that this curve has two phases: the low-temperature magnetized phase and the high-temperature disordered phase, both of which are in the same universality class as pure gravity (i.e., they both have $\gamma_{str} = -\frac{1}{2}$). The boundary between these phases is the critical point of the spins, where their critical fluctuations modify the global geometry, and at which $\gamma_{str} \neq -\frac{1}{2}$, in general. For very large $q$ the picture is different: there is still a low-temperature phase (where $\gamma_{str} = -\frac{1}{2}$), but this now goes into a high-temperature phase of *branched polymers* (where $\gamma_{str} = +\frac{1}{2}$); at the multicritical point separating the two phase $\gamma_{str} = +\frac{1}{3}$. As $q$ is decreased, the old disordered phase makes its re-appearance[4]; the order of the phases is: magnetized at low temperature; branched polymers at intermediate temperature; and disordered at high temperature. As $q$ decreases the magnetized and the disordered phase move closer and closer to each other until, at some $q = q_c$, they touch: the branched polymer phase is eliminated, and the low-$q$ phase diagram is restored [9].

Here is a summary of the results that I get using various series extrapolation techniques on the 10th order Potts model series. First, I calculate, for different values of $q$, $\gamma_{str}$ on the critical $(g, a)$ curve using Padé approximants—this is plotted as a function of $g$ in Fig. 4. For small $q$ there are two $\gamma_{str} = -\frac{1}{2}$ regions, the magnetized and the disordered, separated by a sharp peak at the phase transition (the height of the peak agrees quite well with KPZ for $q = 2$ and 3). As $q$ gets larger, the peak gets broader (an effect also observed in simulations). For high enough $q$, moreover, a plateau seems to appear between the two regions, its level approaching $\gamma_{str} = +\frac{1}{2}$ in the limit $q \to \infty$. To accurately locate the phase transition for $q > 4$ (where it may no longer correspond to the $\gamma_{str}$ peak), and to include logarithmic corrections, I have used a variation on the ratio method. The results are summarized in Fig. 5 and in Table 3. The logarithmic corrections give a spectacular improvement in $\gamma_{str}$ at and just above $q = 4$. For example, for $q = 4$ (where KPZ gives $\gamma_{str} = 0$), I get $\gamma_{str} \approx -0.2$ without logarithmic corrections, but $\gamma_{str} \approx 0.003$ with logarithms. There is a region, approximately $4 \leq q < 15$, where $\gamma_{str}$ increases gradually, while the logarithmic exponent remains roughly constant. Then, for higher $q$, the logarithmic exponent and the corresponding $\gamma_{str}$ decrease, probably indicating that logarithmic extrapolations can no longer be trusted in this regime (where the logarith-mic corrections are small or zero). Above this uncertain region (roughly $15 < q < 50$), it seems clear that one should use non-logarithmic fits. The resulting $\gamma_{str}$ increases slowly and smoothly with $q$, approaching 0.37 in the limit $q \to \infty$ (the exact answer is $\gamma_{str} = +\frac{1}{3}$).

[4] In the language of [9] this is called the phase of "large galaxies."

The rest of this paper is organized as follows. Section 2 is a careful discussion of the LTE (for arbitrary target spaces) and how to calculate the coefficients. Section 3 is a numerical analysis of the Potts example. Conclusions and prospects for future work are all to be found in Section 4.

## 2 Generating the series

This section is a didactic presentation of the mechanics of the LTE. I have tried to give motivations and intuitive explanations of the different factors that enter. Table 1 summarizes the calculations though fourth order. There are two types of factors: ones that do *not* depend on the matter model that is coupled to the surface (sections 2.1–2.4; all but the right-hand column of Table 1); and one that *does* (section 2.5; the right-hand column of Table 1).

### 2.1 Spin clusters

The low-temperature expansion for discretized random surfaces is similar to the LTE for ordinary spin models. It is essentially a spin cluster expansion: in each order $a^n$ (recall that the matter coupling constant $a = (q-1)e^{-2\beta}$—see eqs. (2) and (3)) one sums all configurations that have $n$ links that connect unequal spins. It is convenient to represent these configurations by *skeleton graphs* (or "skeletons" for short). The vertices of a skeleton are the *spin clusters*, that is, connected sets of lattice sites that have a constant spin; while the edges of a skeleton are only those edges of the lattice that connect the spin clusters to each other. The spin clusters are surfaces of arbitrary size (whose fluctuations will be summed by appropriate vertex factors), but with restricted topology. (The reader should not confuse the discretized random surfaces—*i.e.*, the fatgraphs of a matrix model—with the skeleton graphs of the LTE. A skeleton represents a fatgraph with a particular configuration of spins; many such fatgraph/spin configurations can be represented by one skeleton.) By definition, a vertex of a skeleton cannot be connected to itself.

For instance, the skeletons that enter into orders 0–4 of the LTE are given in the left column of Table 1. Weighted by different embedding factors, these graphs should occur in the LTE of any model—random surface or fixed lattice. For any model on a fixed, 2d square lattice however, only two of

| Graph | Sym. | V.F. | Spin cluster/topology | Total | Potts coloring |
|---|---|---|---|---|---|
|  | 1 | 1 | $\pi_0$ | $\rho_0$ | 1 |
|  | $\frac{1}{2}$ | 1 | $\pi_1^2$ | $\frac{1}{2}\rho_1^2$ | $\Delta a$ |
|  | $\frac{1}{2}$ | $\frac{1}{2}$ | $2\pi_1^2(\pi_2+\pi_{1,1})$ | $\frac{1}{2}\rho_1^2\rho_{1,1}$ | $\Delta^2 a^2$ |
|  | $\frac{1}{2}$ | $\frac{1}{4}$ | $2\pi_2^2$ | $\frac{1}{4}\rho_2^2$ | $\Delta a^2$ |
|  | $\frac{1}{2}$ | $\frac{1}{4}$ | $4\pi_1^2(\pi_2+\pi_{1,1})^2$ | $\frac{1}{2}\rho_1^2\rho_{1,1}^2$ | $\Delta^3 a^3$ |
| 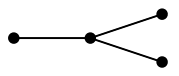 | $\frac{1}{6}$ | $\frac{1}{6}$ | $6\pi_1^2(\pi_3+\pi_{2,1}+\pi_{1,1,1})$ | $\frac{1}{6}\rho_1^3\rho_{1,1,1}$ | $\Delta^3 a^3$ |
| 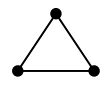 | $\frac{1}{6}$ | $\frac{1}{8}$ | $8\pi_2^3$ | $\frac{1}{6}\rho_2^3$ | $\Delta(\Delta-1)a^3$ |
|  | 1 | $\frac{1}{6}$ | $\pi_1\pi_2(3\pi_3+\pi_{2,1})$ | $\frac{1}{2}\rho_1\rho_2\rho_{2,1}$ | $\Delta^2 a^3$ |
| 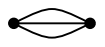 | $\frac{1}{2}$ | $\frac{1}{36}$ | $3\pi_3^2$ | $\frac{1}{24}\rho_3^2$ | $\Delta a^3$ |
|  | $\frac{1}{2}$ | $\frac{1}{8}$ | $8\pi_1^2(\pi_2+\pi_{1,1})^3$ | $\frac{1}{2}\rho_1^2\rho_{1,1}^3$ | $\Delta^4 a^4$ |
| 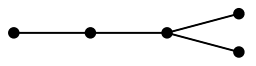 | $\frac{1}{2}$ | $\frac{1}{12}$ | $12\pi_1^3(\pi_2+\pi_{1,1})(\pi_3+\pi_{2,1}+\pi_{1,1,1})$ | $\frac{1}{2}\rho_1^3\rho_{1,1}\rho_{1,1,1}$ | $\Delta^4 a^4$ |
| 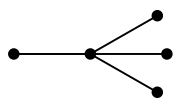 | $\frac{1}{24}$ | $\frac{1}{24}$ | $24\pi_1^4(\pi_4+\pi_{3,1}+\pi_{2,2}+\pi_{2,1,1}+\pi_{1,1,1,1})$ | $\frac{1}{24}\rho_1^4\rho_{1,1,1,1}$ | $\Delta^4 a^4$ |
| 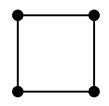 | $\frac{1}{8}$ | $\frac{1}{8}$ | $8\pi_2^2$ | $\frac{1}{8}\rho_2^4$ | $\Delta(\Delta^2-\Delta+1)a^4$ |
| 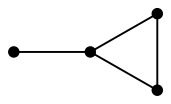 | $\frac{1}{2}$ | $\frac{1}{24}$ | $8\pi_1\pi_2^2(3\pi_3+\pi_{2,1})$ | $\frac{1}{2}\rho_1\rho_2^2\rho_{2,1}$ | $\Delta^2(\Delta-1)a^4$ |
|  | 1 | $\frac{1}{24}$ | $4\pi_1\pi_2(\pi_2+\pi_{1,1})(3\pi_3+\pi_{2,1})$ | $\frac{1}{2}\rho_1\rho_2\rho_{1,1}\rho_{2,1}$ | $\Delta^3 a^4$ |
|  | $\frac{1}{2}$ | $\frac{1}{36}$ | $2\pi_1^2(3\pi_3+\pi_{2,1})^2$ | $\frac{1}{4}\rho_1^2\rho_{2,1}^2$ | $\Delta^3 a^4$ |
| 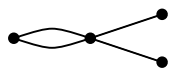 | $\frac{1}{2}$ | $\frac{1}{48}$ | $2\pi_1^2\pi^2(12\pi_4+6\pi_{3,1}+4\pi_{2,2}+2\pi_{2,1,1})$ | $\frac{1}{4}\rho_1^2\rho_2\rho_{2,1,1}$ | $\Delta^3 a^4$ |
| 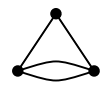 | $\frac{1}{2}$ | $\frac{1}{72}$ | $18\pi_2\pi_3^2$ | $\frac{1}{8}\rho_2\rho_3^2$ | $\Delta(\Delta-1)a^4$ |
|  | $\frac{1}{2}$ | $\frac{1}{96}$ | $4\pi_2^2(4\pi_4+2\pi_{2,2})$ | $\frac{1}{12}\rho_2^2\rho_{2,2}$ | $\Delta^2 a^4$ |
| 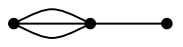 | 1 | $\frac{1}{144}$ | $3\pi_1\pi_3(4\pi_4+\pi_{3,1})$ | $\frac{1}{12}\rho_1\rho_3\rho_{3,1}$ | $\Delta^2 a^4$ |
| 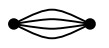 | $\frac{1}{2}$ | $\frac{1}{576}$ | $4\pi_4^2$ | $\frac{1}{288}\rho_4^2$ | $\Delta a^4$ |

Table 1: Graphs through order 4 of the LTE and the factors that they contribute

the graphs in Table 1 occur: • and 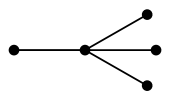 (for example,  does not occur because it is impossible to divide the lattice into two clusters that are connected by just one edge). In fixed-lattice models, each skeleton typically gives rise to three factors: symmetry, embedding, and “homomorphism coloring” (the number of ways to assign spins to the clusters).

For a random surface model, almost all of the graphs in Table 1 will typically contribute. Each skeleton contributes two of the above factors: symmetry and coloring. Now, however, we have a random geometry model. Therefore the spin clusters—the vertices of the skeleton graph—are fluctuating random surfaces, and the contribution from each skeleton must include sums over these fluctuations. These will appear as vertex factors, denoted as $\pi$'s or $\rho$'s in what follows. Each cluster will be a punctured surface, with as many punctures as the degree of the corresponding skeleton graph vertex. Since the spins on each cluster are frozen, the clusters' geometrical fluctuations will be governed by a pure gravity one-matrix model. Finally, there is no restriction on the *area* of the spin clusters: each order of the LTE includes surfaces of all sizes.

Notice that only the coloring factors for the skeletons depend on the details of the matter model that is coupled to the surface. I will return to the coloring factors in section 2.5, while the next sections will be devoted to the topology and geometry of the spin clusters.

## 2.2 Twist and topology

Exactly which surfaces must be summed over for each spin cluster? The answer depends, unfortunately, on the details of each skeleton graph. I will first introduce a classification for *punctured* discretized random surfaces, and then explain which of these contribute to any cluster in any given skeleton.

Closed discretized random surfaces are classified by their genus, or the minimum number of handles. The additional complication will be that spin clusters, since they are (usually) connected to other spin clusters, are open, punctured, surfaces. In the context of the LTE, a natural way to classify punctured surfaces will make use of two parameters: the *genus* and the *twist*. The genus is defined by putting endcaps on the punctures, so that none of the endcaps is connected to any other (see Fig. 1a). The genus of the punctured surface will be defined as the genus of the endcapped version. Our clusters will all be genus-0 surfaces according to this definition.

To define the twist, put a dummy matrix $\Lambda$ on each puncture. The propagators will connect the $\Lambda$'s in some way; the result of these contractions will be (see Fig. 1) $\mathrm{Tr}\,\Lambda^{k_1}\mathrm{Tr}\,\Lambda^{k_2}\cdots$, so that $k_1+k_2+\cdots=k$, where $k$ is the number of punctures; the set $\{k_1,k_2,\ldots\}$ is therefore an (integer) partition of $k$.[5] For a surface with $k$ punctures, the quantity which I will call "twist" will be the corresponding partition $p$ chosen from $P_k$, the set of partitions of $k$; for example, $P_3=\{\{3\},\{2,1\},\{1,1,1\}\}$ (see Fig. 1b). Fig. 2 shows the simplest concrete example of twist, for $k=2$. I will denote the corresponding Green's functions by $\pi_p(g)$, so that the three possibilities for $k=3$ in Fig. 1b will be called (from top to bottom) $\pi_3(g)$, $\pi_{2,1}(g)$, and $\pi_{1,1,1}(g)$.

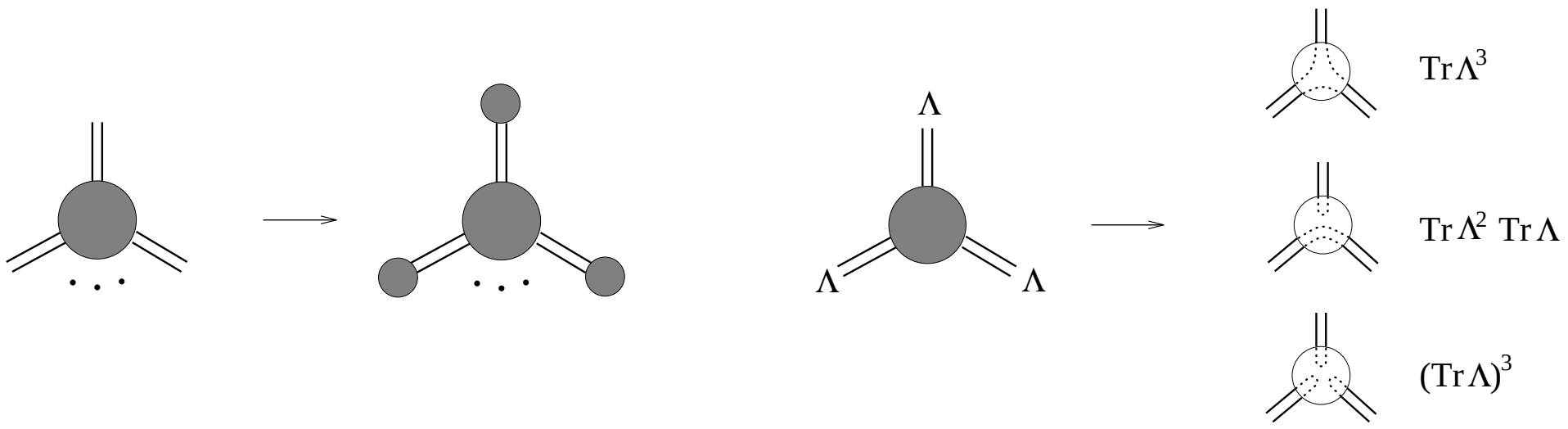

(a) Definition of genus (b) Definition of twist

Figure 1: Classification of punctured surfaces

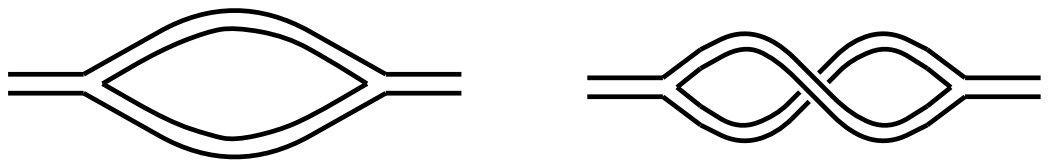

(a) (b)

Figure 2: An example of (a) an untwisted spin cluster (counted by $\pi_2$) and (b) a twisted spin cluster (counted by $\pi_{1,1}$) with two punctures

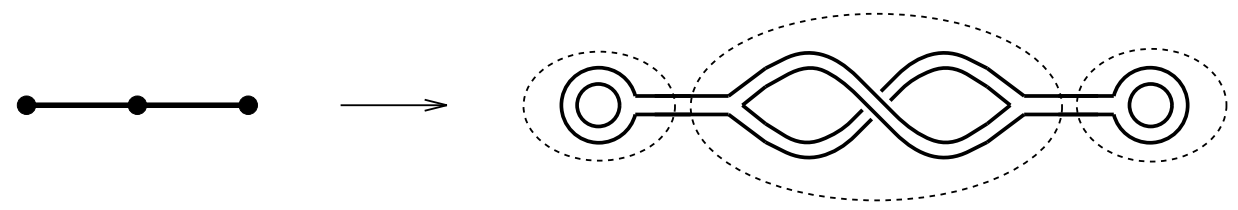

Figure 3: A planar surface with a twisted cluster. The skeleton graph is shown on the left, an example of a possible corresponding surface (fatgraph) on the right.

For a cluster with $k$ punctures, call the surfaces counted by $\pi_k$ *untwisted*, and all the others (counted by $\pi_{k-1,1},\ldots$) *twisted*. The untwisted Green's functions $\pi_k$ are usually called just the "Green's functions"; the twisted surfaces, to my knowledge, have never been considered or calculated (see section 2.3 for how it's done). Your intuition tells you that twisted surfaces cannot contribute to the planar limit; this intuition is false. To properly represent multi-matrix models in terms of clusters, we *must* include *some* twisted clusters. The reason is that, in some cases, twists can be untwisted: the resulting surface is really planar, and therefore must be included. Fig. 3 provides a simple illustration.

Now we can return to question of which skeleton graphs contribute to the partition function. Since we want to have a planar surface in the end, the skeleton itself should be thought of as a fatgraph. Each vertex (spin cluster) with $k$ legs contributes a factor of $k!^{-1}\pi_p(g)$, where $p$ is a twist, i.e., a partition of $k$: this vertex factor is the contribution of the fluctuating geometry of the spin cluster. Thus, $k$-leg vertices come in $\varpi_k$ flavors, where $\varpi_k$ is the number of integer partitions of $k$ (for large $k$, $\varpi_k$ grows as $e^{\sqrt{k}}$). In principle, each $k$-leg vertex contributes a vertex factor which is a sum over all the twists of $k$ with equal weight, $\sum_{p\in P_k}\pi_p$. However, some combinations of twists will result in non-spherical surfaces; so that the net vertex factor is actually some linear combination of the twists, $\sum_{p\in P_k} c_p\pi_p$—see Table 1, column "Spin cluster/topology" for examples.

[5] An *(integer) partition* of $k$ is an orderless set of positive integers whose sum is $k$.

In calculating the linear combinations of twists that occur at each vertex of each skeleton, untwisted clusters $\pi_k$ may be faithfully represented as ordinary linear matrix model vertices ($\mathrm{Tr}\,\Lambda^k$), while the twisted clusters $\pi_{k_1,k_2,\cdots}$ are represented as nonlinear vertices (popularized by [10]) $\mathrm{Tr}\,\Lambda^{k_1}\,\mathrm{Tr}\,\Lambda^{k_2}\cdots$ which appear in models of polymerized surfaces. In other words, a twisted spin cluster is topologically equivalent to a surface touching. This is why the nonlinear vertices occur in the effective potentials of the one-matrix version of the Potts model [11], as well as in the effective potential of the reduced lattice model [12].

The preceding calculation may be simplified by noting that only certain linear combinations of the $\pi$'s occur as vertex factors. To determine which linear combination to take for a given vertex of a given fatgraph, let us define the "connectivity partition" of a vertex. Consider a vertex with $k$ legs in some skeleton. Any two of its legs are said to be connected if one can reach one of the legs starting from the other, *without* going through the vertex itself. The $k$ legs will thus be divided into some number of clusters: every leg is connected to every leg in its cluster, and not connected to any other leg. If the number of legs in the clusters is $q_1, q_2, \ldots$, then the $q$'s are a partition of $k$, the *connectivity partition.* If the graph is tree, for example, for a given vertex no leg will be connected (in the present sense) to any other, and therefore the connectivity partition of every vertex will be $\{1, 1, \ldots\}$. This connectivity partition precisely determines the linear combination of $\pi$'s which will be the required vertex factor. I will denote these vertex factors $\rho_{q_1,q_2,\ldots}$. Here are some examples:

$$
\begin{aligned}
\rho_0 &= \pi_0 \\
\rho_1 &= \pi_1 \\
\rho_2 &= \pi_2 \\
\rho_{1,1} &= \pi_2 + \pi_{1,1} \\
\rho_3 &= \pi_3 \\
\rho_{2,1} &= \pi_3 + \frac{\pi_{2,1}}{3} \\
\rho_{1,1,1} &= \pi_3 + \pi_{2,1} + \pi_{1,1,1} \\
\rho_4 &= \pi_4 \\
\rho_{3,1} &= \pi_4 + \frac{\pi_{3,1}}{4}
\end{aligned}
\tag{4}
$$

$$
\begin{aligned}
\rho_{2,2} &= \pi_4 + \frac{\pi_{2,2}}{2} \\
\rho_{2,1,1} &= \pi_4 + \frac{\pi_{3,1}}{2} + \frac{\pi_{2,2}}{3} + \frac{\pi_{2,1,1}}{6} \\
\rho_{1,1,1,1} &= \pi_4 + \pi_{3,1} + \pi_{2,2} + \pi_{2,1,1} + \pi_{1,1,1,1}
\end{aligned}
$$

The recipe for using the $\rho$ vertex factors is as follows. Instead of summing over all $k$-twists for each $k$-vertex, take only the untwisted vertex, $\mathrm{Tr}\,\varphi^k$. Having calculated the large-$N$ factor using the untwisted vertices, determine the connectivity partition $q$ of each vertex, and multiply by $\rho_q$. Using the $\rho$ rather than $\pi$ vertex factors makes the calculation more transparent and saves one the trouble of summing over all twists for each vertex.

### 2.3 Calculating the clusters

The spin cluster Green's functions $\pi_p$ are generated by an external matrix integral [11, 13]

$$
\Pi(\Lambda) = -\log \int \mathcal{D}\varphi \, \exp - N^2 \mathrm{tr}\,[\Lambda\varphi + V(\varphi)] \tag{5}
$$

where $\Lambda$ is some fixed $N \times N$ hermitian matrix, and as usual matrix integrals are understood in the sense $\lim_{N\to\infty} N^{-2}$ (and $\mathrm{tr}\, X = \mathrm{Tr}\, X/\dim X$). The $\Lambda\varphi$ vertex inserts punctures, and the fact that $\Lambda$ is a matrix keeps track of the twist. Expand the generating function $\Pi(\Lambda)$ in powers of $\Lambda$, one finds precisely

$$
\begin{aligned}
\Pi(\Lambda) &= \pi_0 + \pi_1 \mathrm{tr}\,\Lambda + \frac{1}{2}[\pi_2 \mathrm{tr}\,\Lambda^2 + \pi_{1,1}(\mathrm{tr}\,\Lambda)^2] \\
&\quad + \frac{1}{6}[\pi_3 \mathrm{tr}\,\Lambda^3 + \pi_{2,1}\mathrm{tr}\,\Lambda^2 \mathrm{tr}\,\Lambda + \pi_{1,1,1}(\mathrm{tr}\,\Lambda)^3] + \cdots
\end{aligned} \tag{6}
$$

For the cubic potential $V(\varphi) = \varphi^2/2 + g\varphi^3$ the generating function $\Pi(\Lambda)$ has been calculated exactly [11, 13]. One could expand this result in powers of $\Lambda$ to obtain closed-form expressions for the $\pi_p$ as functions of $g$.

A more efficient method is to use the Schwinger-Dyson equation for integral (5), which reads

$$
\Lambda - N\partial_\Lambda \Pi + 3g[\partial_\Lambda^2 \Pi + N^2(\partial_\Lambda \Pi)^2] = 0 \tag{7}
$$

Inserting the expansion (6) into this equation, one finds in $n$-th order a system of linear equations for the $n$-puncture functions $\pi_p$ ($p \in P_n$), where the coefficients and inhomogeneous terms involve functions with fewer than $n$ punctures; this gives us, therefore, a recursive procedure for calculating the $\pi_p$. The only complication is that the most twisted function in each order ($\pi_{1,\ldots,1}$) does not appear. This is not a problem, though: one can easily calculate the sum over all twists in each order $\tilde{\pi}_n = \sum_{p\in P_n} \pi_p$ by setting $\Lambda = \lambda 1$; eq. (6) now reads

$$\Pi(\lambda) = \pi_0 + \pi_1\lambda + \frac{\tilde{\pi}_2}{2}\lambda^2 + \frac{\tilde{\pi}_3}{6}\lambda^3 + \cdots \tag{8}$$

by simply shifting the matrix in the BIPZ one-matrix integral $\varphi \to \varphi + x1$ by a multiple of unity; this gives us an extra linear equation for $\pi_{1,\ldots,1}$.

The Schwinger-Dyson equations (7) are valid for all $N$, and therefore generate non-spherical corrections; for the sake of efficiency, it would be desirable to eliminate these corrections. This can be done by following a certain prescription for the matrix derivative operators $\partial_\Lambda$ and $\partial_\Lambda^2$. For a partition $p = \{p_1, \ldots, p_k\}$, define $|p| = k$; $p - p_\ell$ as $p$ with the $\ell$-th element removed; and $\chi_p(\Lambda) = \Lambda^{p_1} \cdots \Lambda^{p_k}$. The prescription for the matrix derivatives is

$$\partial_\Lambda \chi_p(\Lambda) = \sum_{r=1}^{|p|} p_r \Lambda^{p_r - 1} \chi_{p-p_r}(\Lambda) \tag{9}$$

$$\partial_\Lambda^2 \chi_p(\Lambda) = \sum_{r=1}^{|p|} p_r \left( \sum_{s=0}^{p_r-2} \Lambda^{p_r-s-2} \,\mathrm{Tr}\,\Lambda^s \right) \chi_{p-p_r}(\Lambda) + \textit{non-spherical} \tag{10}$$

The first derivative (9) is exact; the non-spherical terms omitted in the second derivative (10) are guaranteed to contribute only subleading $\mathcal{O}(N^{-1})$ terms to the Schwinger-Dyson equations; moreover, all the subleading terms in the Schwinger-Dyson equations are eliminated by simply omitting the "non-spherical" terms in eq. (10). Putting eqs. (7), (6), (9) and (10) together, the first few equations of the Schwinger-Dyson hierarchy are

$$3g(\pi_2 + \pi_1^2) - \pi_1 = 0 \tag{11}$$

$$3g\left(\frac{\pi_3}{2} + 2\pi_1\pi_2\right) - \pi_2 + 1 = 0 \tag{12}$$

$$3g\left(\frac{\pi_3}{2} + \frac{\pi_{2,1}}{3} + 2\pi_1\pi_{1,1}\right) - \pi_{1,1} = 0 \tag{13}$$

Except for $\pi_0$, the other $\pi_p$'s are rational functions of $g$ and $\sigma(g)$, which satisfies the equation [14] $18g^2 + \sigma(1+\sigma)(1+2\sigma) = 0$; in other words,

$$\sigma(g) = -\frac{1}{2} + \frac{1}{\sqrt{3}}\cos\frac{1}{3}[\pi - \cos^{-1}(g/g_0)^2] \qquad g_0^2 = \frac{1}{108\sqrt{3}} \tag{14}$$

Starting with [14]

$$\pi_1 = -\frac{\sigma(1+3\sigma)}{6g(1+2\sigma)} \tag{15}$$

one finds

$$\pi_2 = -\frac{\sigma(3+15\sigma+16\sigma^2)}{2(1+2\sigma)^3} \qquad \pi_{1,1} = -\frac{\sigma(1+\sigma)}{2(1+2\sigma)^3} \tag{16}$$

$$\pi_3 = -\frac{\sigma(1+\sigma)(1+5\sigma+5\sigma^2)}{3(1+2\sigma)^4 g} \qquad \pi_{2,1} = \frac{\sigma^2(1+\sigma)^2}{(1+2\sigma)^4 g} \tag{17}$$

$$\pi_{1,1,1} = -\frac{4\sigma^3(1+\sigma)^3}{3(1+2\sigma)^4(1+6\sigma+6s^2)g} \tag{18}$$

and so on.

### 2.4 Punctures

So far I have been discussing how to calculate the LTE of the free energy $F_q(g,a)$. By definition, its singularity is $F_q \approx t^{2-\gamma_{str}}$, where $t$ is the distance from the critical point (for models with multicritical behavior, this definition assumes that we approach the critical point from a generic direction). Now, since $\gamma_{str}$ is expected to lie in the range $[-\frac{1}{2}, +\frac{1}{2}]$, this singularity is rather weak, and consequently difficult to study using series methods.

Fortunately, this problem can be circumvented by *puncturing* the surface. A puncture is a labeled, degree-1 vertex, *either* in the original matrix-model fatgraph, *or* in the skeleton graph. I will denote the connected $n$-puncture Green's function $G_n$ ($F_q = G_0$). The advantage of the punctures is that each one brings the power of the singularity down by one: $G_n \approx t^{2-n-\gamma_{str}}$. As for the matter, one has three options: putting fluctuating spins on the punctures, putting fixed spins, or putting no spins at all. I will take the last option, as it is the simplest. In this case the vertex factor for each puncture in the skeleton graph is just unity.

Making three or more punctures, therefore, ensures that the critical point will always be a divergence. The stronger the divergence, the easier it is to study it using series extrapolation techniques; practically, though, making too many punctures is calculationally cumbersome. As a compromise I have chosen to make four punctures.

Starting with the free energy, in the general case it is of course impossible to obtain the Green's functions. In the LTE representation, however, one can puncture the surface in a completely mechanical way. The prescription is as follows: for every partition $p$, replace the corresponding vertex factor $\pi_p$ in $F_q = G_0$ by the series

$$\pi_p \to \pi_p + \pi_{p,1}x + \frac{1}{2}\pi_{p,1,1}x^2 + \cdots + \frac{1}{k!}\pi_{p,\underbrace{1,\ldots,1}_{k}}x^k + \cdots \qquad (19)$$

Expand everything in powers of $x$: $G_n$ will then simply be the coefficient of $x^n$ divided by $n!$. The reason this works is that the topology of punctures is trivial; puncturing a fatgraph does not change its topology.

## 2.5 Coloring the skeletons

Each skeleton graph in the LTE contributes one last factor: the number of ways to assign a spin or matter state to each spin cluster, i.e., the number of ways to "color" its vertices. This is the only factor in the LTE that depends on the matter model that is coupled to the surface: the matter model defines the coloring rules.

The matter model is defined by the matrix $T_{ij}$ that appears in the definition of the multi-matrix model, eq. (2); the entries of this matrix are functions of $z = e^{-2\beta}$, the matter temperature. $T_{ij}$ can be considered as the adjacency matrix of the target space graph ("target graph" for short), $\mathcal{T}$, each of whose vertices represents a matter state. If $T_{ij} = 0$ then vertices $i$ and $j$ in $\mathcal{T}$ are not connected; otherwise they are, with the weight $T_{ij}$. This is how the coefficients of the LTE depend on the target graph $\mathcal{T}$: each spin cluster (in each skeleton) must be "colored," i.e., assigned a matter state (matter state = vertex of $\mathcal{T}$), but only in such a way as is allowed by $\mathcal{T}$ and weighted by a correponding factor. Each skeleton contributes a sum over all such colorings.

More formally: consider a skeleton graph $\mathcal{S}$. A coloring of $\mathcal{S}$ is a mapping $\tau$ from the vertices of $\mathcal{S}$ to the vertices of $\mathcal{T}$. The coloring factor $\chi(\mathcal{S})$ is a sum over all distinct mappings $\tau$. Each mapping is weighted by a product over the edges of $\mathcal{S}$: an edge between vertices $i$ and $j$ contributes the factor $T_{\tau(i),\tau(j)}$. Symbolically,

$$\chi(\mathcal{S}) = \sum_{\tau:\mathcal{S}\mapsto\mathcal{T}} \prod_{\{i,j\}\in\mathcal{S}} T_{\tau(i),\tau(j)} \qquad (20)$$

In other words, a mapping $\tau$ is allowed only if every pair of nearest neighbors $\{i,j\}$ in $\mathcal{S}$ is mapped to a pair of nearest neighbors $\{\tau(i),\tau(j)\}$ in $\mathcal{T}$. Such mappings are called "graph homomorphisms" or "H-colorings," and were introduced into graph theory by Nešetřil [15].

The simplest matter model in some sense is the $q$-state Potts model. Its target graph $\mathcal{T}$ is the complete graph on $q$ vertices $K_q$, i.e., $q$ vertices, each of which is connected to all the others. Every edge in $\mathcal{T}$ has equal weight, $z$. What are the coloring rules for this model?

Every vertex of $\mathcal{T}$ is connected to every other vertex besides itself: this means that all mappings are allowed *except* ones that assign the *same* color to nearest-neighbor vertices in the skeleton. This is precisely what is meant in graph theory by "properly coloring" the graph $\mathcal{S}$ with $q$ colors. The number of such proper colorings is called the *chromatic polynomial* of $\mathcal{S}$, $\chi_q(\mathcal{S})$: it is a polynomial in $q$ of degree less than or equal to the number of vertices $n$ in $\mathcal{S}$. If S is a tree then the chromatic polynomial is trivially $q(q-1)^{n-1}$; otherwise it is a polynomial of degree less than $n$. To recapitulate: the coloring factors for the Potts model are just the chromatic polynomials. These are given in Table 1 in terms of $\Delta = q - 1$.[6]

I use a simple and elegant algorithm, originally due to Birkhoff, for computing the chromatic polynomials. Consider a skeleton graph $\mathcal{S}$ and let $\mathcal{S}'$ be $\mathcal{S}$ with some edge $\{i, j\}$ deleted (provided that $\mathcal{S}'$ is connected). Deleting the edge increases the chromatic polynomial: $\chi_q(\mathcal{S}) < \chi_q(\mathcal{S}')$ for any $q$, since there are fewer restrictions on coloring in $\mathcal{S}'$. The difference is precisely due to those colorings for which $\tau(i) = \tau(j)$, i.e., the colors of $i$ and $j$ coincide; in other words, if we contract $i$ and $j$ to a single point and call the resulting graph $\mathcal{S}''$, we have

$$\chi_q(\mathcal{S}) = \chi_q(\mathcal{S}') - \chi_q(\mathcal{S}'') \tag{21}$$

By applying the reduction (21) recursively, one eventually ends up with graphs that are trees, whose chromatic polynomials are trivially $q(q-1)^{n-1}$ ($n$ is the number of vertices of the tree). An efficient algorithm that uses this method has been implemented by Wilf and Nijenhuis [16].

Given a particular skeleton graph $\mathcal{S}$ and a particular target space graph $\mathcal{T}$, one can always calculate the homomorphism coloring factor from $\mathcal{S}$ onto $\mathcal{T}$. For the simple case of the Potts model (where $\mathcal{T} = K_q$), we can do more: for a given $\mathcal{S}$ we can calculate, using Birkhoff's algorithm, the coloring factor onto $K_q$ for arbitrary $q$—the chromatic polynomial of $\mathcal{S}$. This means that the coefficients of the LTE will be explicit functions of $q$—one does not have to redo the calculation for every new value of $q$, and one can plug in $q$'s that are not positive integers.

[6] Since my convention is to normalize matrix model expectation values by the total "volume" of the target graph, i.e., the number of matter states, the coloring factors in Table 1 are the chromatic polynomials divided by $q$.

Other models, such as *multiple* Ising models, are potentially more interesting (since they yield conformal theories even on the fixed lattice). The target graph $\mathcal{T}$ for $\nu$ Ising models is a $\nu$-dimensional hypercube. Its edges are weighted by $z$, corresponding to changing the state by flipping one spin; since more than one spin can be flipped at a time, however, the vertices of the hypercube are connected along the diagonals, which connections are weighted by higher powers of $z$. The varying weights in this target graph make the problem of calculating the coloring factor more complicated that the Potts case ($\mathcal{T} = K_q$), as the simple Birkhoff reduction formula (21) no longer holds.

### 2.6 Putting it all together

This section summarizes and brings together the calculations required to develop the LTE.

First, we need a list of skeleton graphs. At this stage, we consider skeletons as ordinary, thin graphs. To save computational effort, the skeletons should be *unlabeled*. Initially, we make only one restriction: vertices are not allowed to be connected to themselves (this would contradict the definition of spin clusters); multiple edges between vertices are, on the other hand, allowed.[7] We will group the skeletons by their number of edges: $n$ edges contributes to $n$-th order of the LTE. It is not trivial to generate such unlabeled graphs. The graphs used here have been generated by Brendan McKay using his software package "nauty" [17]. All the skeletons through order 4 are shown in the left-hand column of Table 1. Although there are many fewer unlabeled graphs than labeled ones, the number of unlabeled graphs grows faster than exponentially (there are 1183 skeletons in order 8, 4442 in order 9, and 17,576 in order 10). Since we are using unlabeled skeleton graphs, each graph has a symmetry factor, the usual inverse order of the graph's automorphism group. These have also been calculated by "nauty", and are given in the "Sym." column of Table 1.

Next, we must consider the skeletons themselves as fatgraphs. First off, each vertex with $k$ legs contributes $1/k!$; to avoid confusion, the products of these vertex factorials are listed separately in the "V.F." column of Table 1. Then recall that each vertex comes in several flavors, the twists, and that each twist contributes a different vertex factor: for a vertex with 3 legs, the possibilities are $\pi_3$, $\pi_{2,1}$, and $\pi_{1,1,1}$. Topologically, the twists may be represented as nonlinear vertices: for 3 legs we have $\operatorname{tr}\varphi^3$, $\operatorname{tr}\varphi^2\operatorname{tr}\varphi$, and $(\operatorname{tr}\varphi)^3$. For each vertex in a given skeleton one must sum over all the possible twists. For a given combination of twists (on a given skeleton graph), we then calculate the usual large-$N$ fatgraph factor: assuming that the legs are rigid and labeled, this is the number of ways of connecting the legs so that the fatgraph is spherical. For each skeleton this yields a polynomial in the $\pi$ vertex factors (which are themselves non-polynomial in $g$), given in the "Spin cluster/topology" column of Table 1. The "Total" column gives the product of the symmetry, vertex factorial, and spin cluster/topology factors, expressed using the convenient $\rho$ vertex factors (which are just linear combinations of the $\pi$'s—see section 2.2). Finally, in order to ease the subsequent numerical analysis, the surfaces need to be punctured. This is done using eq. (19)—see section 2.4. In the analysis in the following section, I will make four punctures.

[7] Graph theorists would call these "free, undirected multigraphs without slings."

In order to specialize to a particular target space (matter model), one must calculate a homomorphism coloring factor for each skeleton onto the target space—see section 2.5. For the $q$-state Potts model, this coloring factor is just the ordinary chromatic polynomial. Through order 4, these are given in the right-hand column of Table 1, normalized by $q$ and expressed in terms of $\Delta = q - 1$. (Recall that $a = \Delta z$.)

In the present work I have carried out the LTE for the Potts model through order $a^{10}$. Computationally this has not been very taxing. Using rather elementary algorithms implemented in Fortran, the whole calculation took about *two hours* of computer time on an HP 9000/700 workstation. (Actually plugging in various values of $g$ and $q$ into the expressions obtained for the coefficients also takes a non-negligible amount of time.)

The resulting series have been checked in two ways. One can set $q = 2$ or $q = \infty$ and compare to a low-temperature expansion of the two available exact solutions. Alternatively, one can carry out, by hand, an expansion in powers of $g$ for any $q$ (the so-called strong-coupling expansion) up to order $g^m$ (the coefficients will be polynomials in $a$); grouping together powers of $a$, we obtain an approximate LTE to order $a^{3m/2}$ (approximate because each coefficient—a non-polynomial function of $g$—is given as a Taylor expansion to order $g^m$).

# 3 Example: the Potts model

## 3.1 Preliminaries

Series extrapolation is always a tricky business, and especially in the present case, where each coefficient of $a^n$ is itself a complicated function of $g$ and $q$. Fortunately, there are several techniques available (the Padé and ratio methods, and countless variations thereon), which allows for cross-checking; all results to be presented are supported by more than one technique. In addition, much is known exactly for the cases $q \leq 4$ and $q = \infty$, which also provides a valuable check.

Near its critical point $a_c(q,g)$, I will assume that the 4-puncture function has the form

$$G_4(q,g,a) \approx |a - a_c(q,g)|^{-2-\gamma_{str}(q,g)} (\log |a - a_c(q,g)|)^{\lambda(q,g)} \tag{22}$$

(up to multiplicative and additive analytical corrections). The only reason to assume this form is that it is correct for the exactly solved cases ($q < 4$ and $q = \infty$). The form of the logarithmic correction is really a guess; it is exact for the $d = 1$ model, where $\gamma_{str} = 0$ and $\lambda = -1$. Expanding in powers of $a$

$$G_4(q,g,a) = \sum_n c_n(q,g) a^n, \tag{23}$$

one finds asymptotically for high orders

$$c_n(q,g) \sim a_c(q,g)^{-n} n^{1+\gamma_{str}(q,g)} (\log n)^{\lambda(q,g)} \tag{24}$$

For a given $q$, there is a critical curve $(g, a_c(q,g))$, on which the surfaces become large (the continuum limit); on this curve there might be one (or more) multicritical points $(g_*(q), a_*(q))$ which correspond to the phase transition(s) of the matter. A quantity that will be useful is the free energy density of the matter in the limit of large surfaces. This can be obtained as follows: first rewrite (23) as a series in $g$:

$$G_4(q,g,a) = \sum_n d_n(q,a) g^n \tag{25}$$

with the asymptotics of the coefficients

$$d_n(q,a) \sim g_c(q,a)^{-n} n^{1+\gamma_{str}(q,a)} (\log n)^{\lambda(q,a)} \tag{26}$$

The $n$-th term of the strong-coupling expansion (25) counts surfaces of area $n$: the partition function of the matter in the limit of large area ($n \to \infty$) is then precisely $1/g_c(q,a)$, which can be obtained by inverting the function $a_c(q,g)$, calculated using (24). The free energy of the matter,

$$U(q,\beta) = -\log g_c(q,a), \quad \beta = -\log a/(q-1) \tag{27}$$

should have the following scaling form near a multicritical point $(g_*, \beta_*)$

$$U(q,\beta) \sim |\beta - \beta_*|^{2-\alpha(q)} \tag{28}$$

It is important to distinguish the $G_4$ scaling relation (22) from the $U$ scaling relation (28). In the first we approach the critical $(g,a)$ curve from some generic direction; while in the second we are constrained to that curve, and approach the multicritical point $(g_*, a_*)$. The $U$ scaling may also have logarithmic corrections (as for the $q \to 1$ case, where $U \sim (\beta_* - \beta)^2 \log(\beta_* - \beta)$).

The crucial problem is locating the phase transition. First of all, for large $q$, there are probably two transitions rather than one: between the low-temperature magnetized phase and the branched polymer phase; and between the branched polymer phase and the high-temperature disordered phase. The high-temperature transition seems rather hard to detect, however. Both in the LTE and in Monte Carlo experiments [18], the transition from the branched polymer phase to the disordered phase seems gradual, though there are some very good theoretical reasons that it *not* be gradual [19]. Most likely, the finite-size effects smooth out the high-temperature transition much more than the low-temperature one; though why this should be so is a mystery. In any case, the phase transition that I will discuss in the rest of this paper will be the low-temperature one: the one *out of* the magnetized phase, whether it be into the disordered phase (for low $q$), or into the branched polymer phase (high $q$).

I will use two methods to locate the phase transition. First, for $q \leq 4$, I will assume that there are only two phases, the magnetized and the disordered (both with $\gamma_{str} = -\frac{1}{2}$), and at the phase transition $\gamma_{str} > -\frac{1}{2}$. So the way to locate the phase transition, as well as to calculate $\gamma_{str}$, is to find the maximum of $\gamma_{str}$ along the critical curve (the continuum limit). This method is used in conjunction with Padé approximants—see section 3.2.

For larger $q$ this method is unreliable, however, due to the probable presence of the branched polymer phase (which should have $\gamma_{str} = +\frac{1}{2}$) between the ordered and the disordered phases (which still have $\gamma_{str} = -\frac{1}{2}$). Therefore the exponent at the phase transition $-\frac{1}{2} < \gamma_{str} < +\frac{1}{2}$ is no longer the maximum $\gamma_{str}$. Another way to find the phase transition is to use the peaks of the of the derivatives of $U$, $U_k \equiv \partial^k U/\partial\beta^k$; this is used in conjunction with the ratio method—see section 3.3. The peak in the specific heat, $U_2$—and in all higher even derivatives, $U_4, U_6, \ldots$—should indicate the phase transition, with one exception. Assume that the exponent $\alpha$ is an integer, $n$, and that there are *no* logarithmic corrections to the scaling relation, eq. (28). This is indeed the case for $q = 2$ and $q = \infty$, where $\alpha = -1$. This means that $U, U_1, \ldots, U_{1-n}$ are continuous at the phase transition, and the slope on either side is finite; the higher derivatives $U_{2-n}, \ldots$ are discontinuous at the critical point, but again with finite slopes. Take $\alpha = -1$, for example: although the specific heat $U_2$ might still have a peak, this is not necessarily true, and will not be the case if the slopes on either side of the critical point have the same sign. The next derivative, $U_3$, will have a discontinuity, but since there are no divergences this does not have to result in a peak, or it might give *two* peaks, one on either side of the transition. It is only $U_4$ that must have a peak roughly at the critical point. In general, for $\alpha = n$ with no logarithmic corrections, $U_{3-n}, U_{5-n}, \ldots$ should have single peaks near the critical point, while $U_{2-n}, U_{4-n}, \ldots$ should have double peaks, one on either side of the critical point.

To summarize: if $\alpha$ is not an integer, or if it is an integer but with logarithmic corrections, then $U_2, U_4, \ldots$ should have peaks roughly at the critical point. If $\alpha$ is some integer $n$ but *without* logarithmic corrections, we should look for peaks in $U_{3-n}, U_{5-n}, \ldots$ to indicate the phase transition. What is the likely range of $\alpha$? Between $q = 2$ and $q = \infty$ I will assume that $\alpha \geq -1$, the endpoint values. We know from KPZ scaling that $\alpha(q = 4) = 0$ (it is not known whether there should be logarithmic corrections). From a Monte Carlo study [18] we know that at $\alpha(q = 10)$ is very close to 0, while at $q = 200$ it is down to roughly $-0.85$. It seems that in all plausible cases, therefore, $U_4$ should have a peak near the critical point, *except if* $\alpha = 0$ *without logarithmic corrections*, in which case the $U_3$ peak should be a better indicator of the critical point. Not wishing to make *a priori* assumptions about $\alpha(q)$, I will therefore present two candidates for the critical point, the locations of the $U_3$ peak and the $U_4$ peak. It will be seen that the $U_4$ peak fits well for small and large $q$, while the $U_3$ peak makes more sense for an intermediate range of $q \geq 4$, as expected.

### 3.2 Padé approximations

The Padé method [20] assumes that $\lambda = 0$. If there are, indeed, logarithmic corrections, what one obtains for $\gamma_{str}$ is an effective exponent which is shifted from the true value in the same direction as the sign of $\lambda$. One fits the logarithmic derivative of (22)

$$\frac{\partial}{\partial a} \log G_4(q, g, a) \approx \frac{-2 - \gamma_{str}}{|a - a_c|} \tag{29}$$

by a rational function, degree-$L$ in the numerator and degree-$M$ in the denominator

$$[L, M] = \frac{p_0 + p_1 a + p_2 a^2 + \cdots + p_L a^L}{1 + q_1 a + q_2 a^2 + \cdots + q_M a^M} \tag{30}$$

The coefficients can be calculated by simply expanding the right-hand side of (30) in powers of $a$, and equating to the logarithmic derivative of $G_4$ (though there are more effecient methods). Therefore, if the series is known through order $a^n$, once can calculate all the $[L, M]$ provided that $L + M \leq n - 1$. The zeros of the denominator of (30) give the locations of the singularities (some of which are physical and some of which are spurious), and their residues give the critical exponents. By constructing a number of different $[L, M]$ approximants, one can get a sense for which are the physical poles (the ones that recur), and a subjective impression of the error. As is often the case, the $[L, L-1]$ and $[L, L]$ approximants seem to be most accurate.

As an example of the Padé method, Fig. 4 shows the $[5, 4]$ approximant for the exponent $\gamma_{str}$ for the physical pole as a function of the cosmological constant $g$, for $q = 2, 4, 10, 25, 100, 500,$ and 10000 (as $q$ gets larger, the peaks get higher). The right-hand side of the graph corresponds to low temperature and pure gravity: the magnetized phase, hence $\gamma_{str} = -\frac{1}{2}$. The curves were cut off on the left when $a_c$ reaches $\Delta = q - 1$, i.e., when the matter temperature reaches infinity. For a given value of $q$, as one varies $g$ the locations of the physical and spurious poles vary. When a spurious pole crosses the physical pole, one sees a little "glitch": a localized divergence in $\gamma_{str}$ as a function of $g$. In Fig. 4, such glitches have been smoothed over to make the plots easier to read.

For low $q$, certainly for $q \leq 4$, we can interpret the curves as follows: the flat, $\gamma_{str} \approx -\frac{1}{2}$ regions on the right- and left-hand sides of the peak correspond to the pure gravity, magnetized and disordered regimes, respectively; while the peak corresponds to the phase transition, at which the divergent correlation length of the spins (for $q \leq 4$, at least) modifies the global geometry of the surface and therefore $\gamma_{str}$ (or alternatively, the point at which the flat conformal field theory is dressed by gravity). According to the KPZ

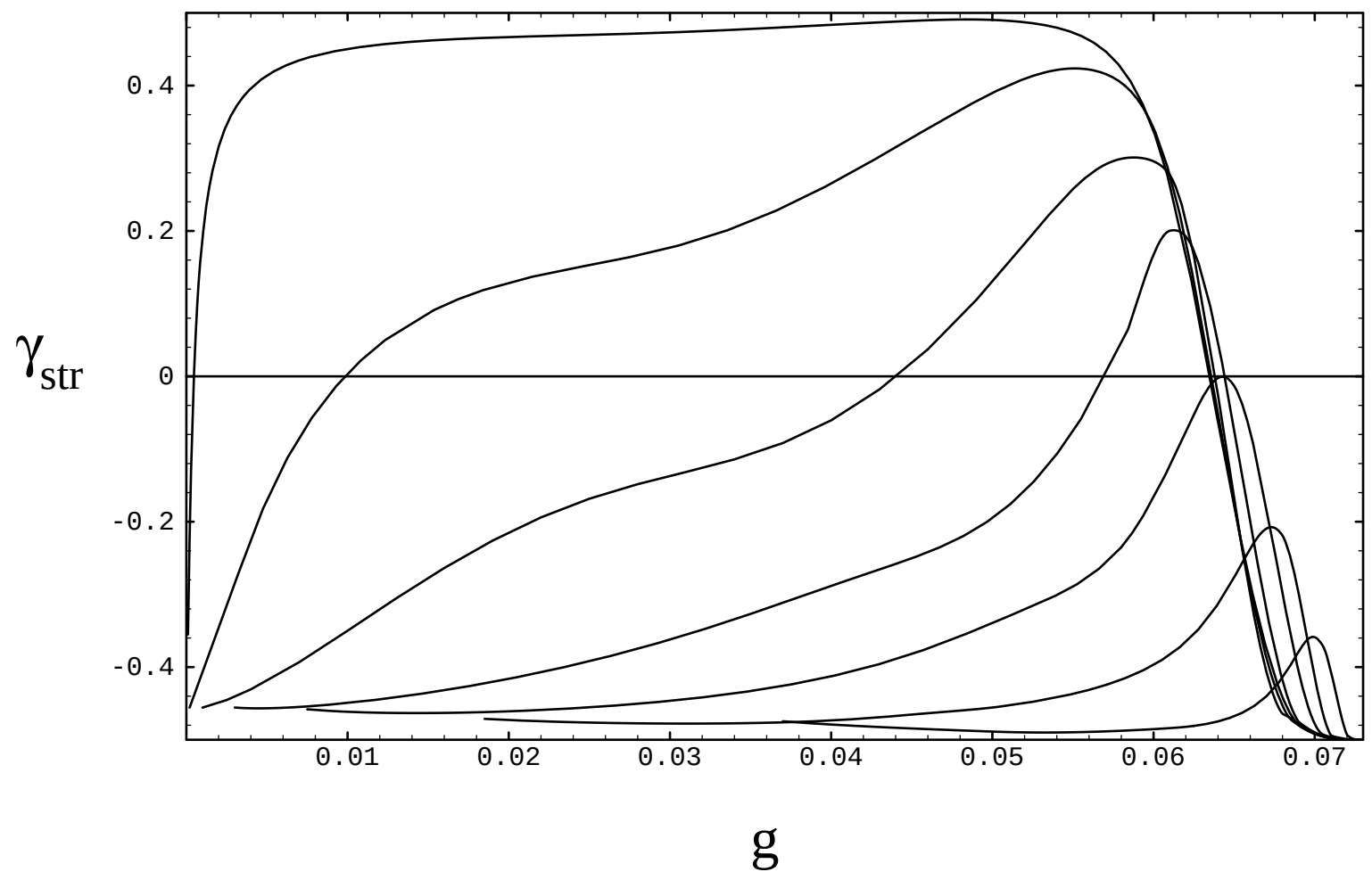

Figure 4: [5,4] Padé estimate of $\gamma_{str}$ as a function of the cosmological constant $g$, for different values of $q$ (in order of ascending peaks, $q = 2, 4, 10, 25, 100, 500, 10000$)

| $q$ | Exact | | | [4,3] Padé | | | [5,4] Padé | | |
|---|---|---|---|---|---|---|---|---|---|
| | $g_c$ | $a_c$ | $\gamma_{str}$ | $g_c$ | $a_c$ | $\gamma_{str}$ | $g_c$ | $a_c$ | $\gamma_{str}$ |
| 2 | 0.0694 | 0.159 | $-1/3$ | 0.0703 | 0.135 | $-0.371$ | 0.0699 | 0.146 | $-0.358$ |
| 3 | 0.0677 | 0.254 | $-1/5$ | 0.0687 | 0.222 | $-0.286$ | 0.0682 | 0.240 | $-0.266$ |
| 4 | | | 0 | 0.0672 | 0.303 | $-0.204$ | 0.0673 | 0.300 | $-0.207$ |

Table 2: Results from Padé analysis for low $q$ using $\gamma_{str}$ peaks to locate the phase transition

formula, valid for $q \leq 4$, the value of $\gamma_{str}$ at the transition is

$$\gamma_{str} = \left(1 - \frac{\pi}{\cos^{-1}\sqrt{q}/2}\right)^{-1} \tag{31}$$

The results for low values of $q$ are presented in Table 2 (the exact $a_c$ and $g_c$ are taken from [21, 22]). The results are quite different for $q = 2$ and 3 than for $q = 4$. While the location of the critical point and $\gamma_{str}$ are not very accurate for $q = 2$ and 3, they are clearly moving in the right direction as one goes from the 8th order of the LTE (the [4,3] Padé) to the 10th order (the [5,4])[8]. For $q = 4$, on the other hand, the results hardly change at all from 8th to 10th order; $\gamma_{str}$ actually gets a little worse. This "sluggishness" is symptomatic of the following general phenomenon: when there logarithmic corrections to scaling are present the convergence of Padé (and other) approximants is drastically slowed. It should also be noted that a value of $\gamma_{str} \approx -0.2$ is commonly obtained in Monte Carlo simulations for various $c = 1$ models.

In Fig. 4 one clearly sees a broadening of the $\gamma_{str}$ peak, as has been observed in Monte Carlo experiments [23] and in strong-coupling series study [7]. For the larger $q$, what we have no longer looks like a peak, but rather a plateau. Could this be the branched polymer phase, predicted in [9] to lie between the low-temperature (magnetized) and the high-temperature (disordered) phases? The presence of a plateau (which extends all the way to $g = 0$ for $q = \infty$) for large $q$ means that the "peak" in $\gamma_{str}$ can no longer be used as a way to detect the phase transition. Indeed, the "peak" values of $\gamma_{str}$ clearly approach $+\frac{1}{2}$ as $q \to \infty$ (the branched polymer value), while we know that *at* the phase transition, $\gamma_{str} = \frac{1}{3}$ when $q$ is sufficiently large. In what follows, I will use peaks in the specific heat and its derivatives to locate the phase transition.

[8]Because of the logarithmic derivative (29), given the series to order $a^n$ (including the zeroth order term), we can calculate Padé approximants up to $L + M = n - 1$

Although it gives the correct general picture, there are three reasons why the Padé method is ultimately not very useful. First, it is difficult to modify it to include logarithmic corrections to scaling ($\lambda \neq 0$). Second, there is no general procedure to extrapolate from finite $[L, M]$ to $L, M \to \infty$. And finally, as the glitches due to root crossings have a strong and random effect on $a_c$ (and a stronger effect on its derivatives), one cannot reliably use the Padé method to look for peaks in the specific heat (and its derivatives) in order to locate the phase transition for high $q$.

### 3.3 Ratio method

A more adequate procedure in this case is the ratio method [20]. The first step is to take the ratios of the coefficients $c_n(q,g)$ in (23), which should have the asymptotics (24):

$$r_n = \frac{c_n}{c_{n-1}} \sim \frac{1}{a_c} \left(\frac{n}{n-1}\right)^{1+\gamma_{str}} \left(\frac{\log n}{\log(n-1)}\right)^{\lambda} \qquad (32)$$

Typically, one expands the right-hand side of (32) in inverse powers of $n$, the constant term giving $a_c$, the coefficient of $1/n$ giving $\gamma_{str}$, and the coefficient of $1/(n \log n)$ giving $\lambda$; these can then be extrapolated to $n \to \infty$, their extrapolants further extrapolated, etc., giving successively more refined estimates. This procedure works for the LTE, but I will adopt a slightly different method which seems to be more stable and, for the cases where the answers are known, more accurate.

For each value of $n$ (and $q$ and $g$), calculate the "running" values of $a_c(n;q,g)$, $\gamma_{str}(n;q,g)$, and $\lambda(n;q,g)$ by solving eq. (32) for $n$, $n-1$, and $n-2$. Assuming the scaling form (22), with multiplicative and additive corrections that are analytic at $a_c(q,g)$, the running values should approach their $n \to \infty$ limits with corrections that are $\mathcal{O}(1/n)$; and, if logarithms are present, with additional corrections of $\mathcal{O}(1/n \log n)$. Fitting the running values to the above form, one estimates the asymptotic values.

The are two main questions to be answered: are there logarithmic corrections to scaling, and where exactly is the critical point? If logarithms are indeed present, we may safely fit to the asymptotic form (32). If not, however, fitting to logarithms can be quite dangerous. We know that there are no logarithms for $q < 4$ and $q = \infty$; and indeed, if one fits to (32) there, one gets nonsense. So the question is, for which values in the range $4 \leq q < \infty$ should one fit to logarithms? In what follows, I will hedge my bets, fitting to (32), as well as to the same form with $\lambda = 0$. I will locate the critical point using peaks in $U_3$ and $U_4$. As discussed above, the $U_3$ peak should be used only if $a = 0$ without logarithmic corrections, and otherwise the $U_4$ peak should be used. I will present data at both peaks.

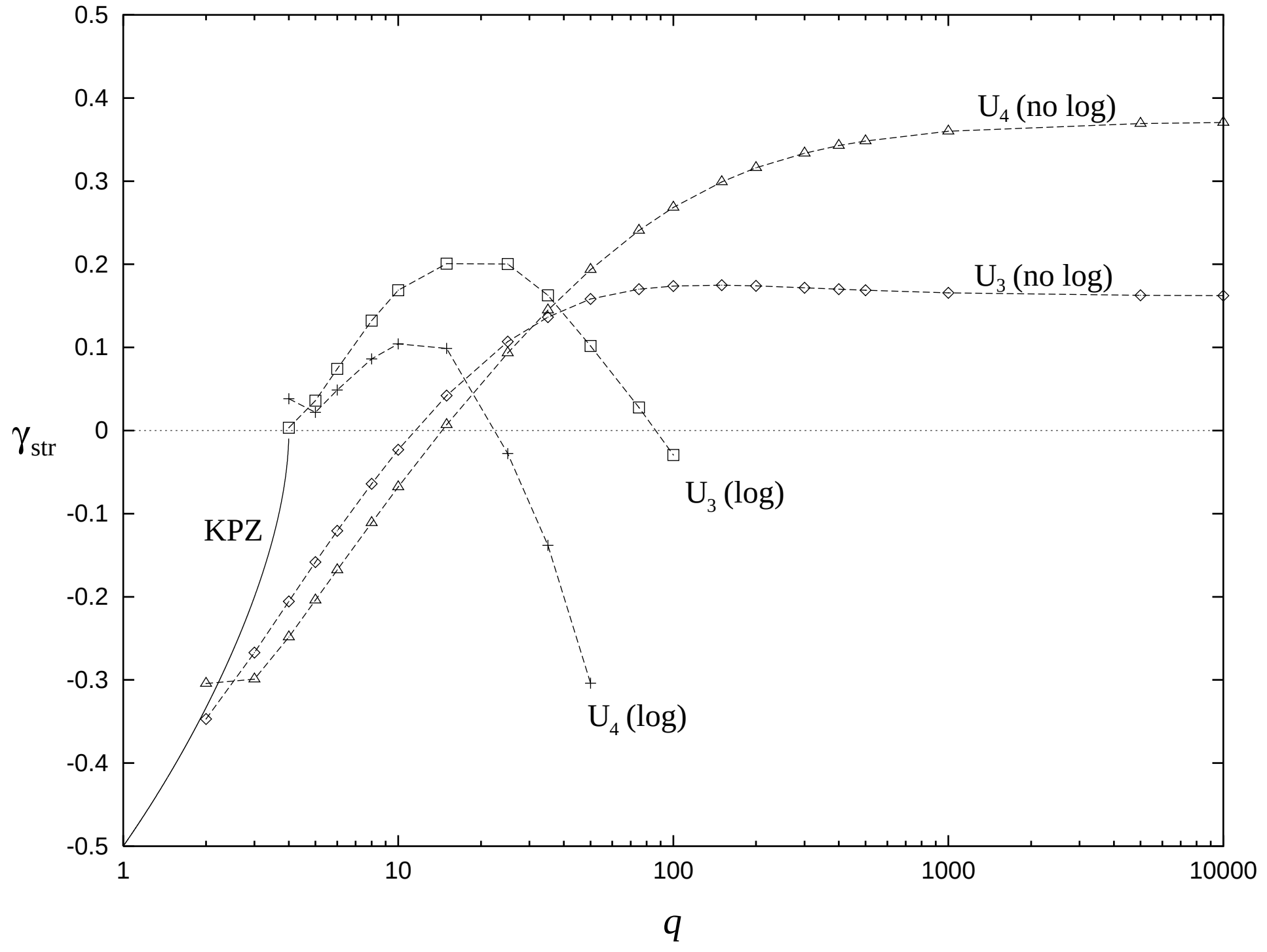

Figure 5: Values of $\gamma_{str}$ obtained as a function of $q$, using different methods to locate the critical point, with and without logarithmic scaling (see text). The KPZ values are also shown for $q \leq 4$.

Fig. 5 presents values of $\gamma_{str}$ calculated for a wide range of $q$ using the methods discussed above. The phase transition is located using the peak in $U_3$ or $U_4$; and the data are fitted with and without logarithmic corrections, For some values of $q$, more complete data are given in Table 3.

There seem to be three different regimes: low $q$ ($q < 4$), intermediate $q$ ($4 \leq q < 50$, rougly), and high $q$ ($50 < q \leq \infty$). First, low $q$. According to KPZ, $\alpha = -1$ (discontinuity) for $q = 2$ and $\alpha = -\frac{1}{2}$ for $q = 3$ (so we should use the peaks in $U_4$ for locating these transitions), and there are no logarithms (i.e., $\lambda = 0$). For $q = 2$, the $U_4$ peak gives a very good value of $a_c = 0.1586$, compared to the exact answer, $a_c = 0.1589$, as well as a decent value for $\gamma_{str}$. For $q = 3$, $a_c$ is still very close to the true value $a_c = 0.254$; $\gamma_{str}$, on the other hand, is not very good. Part of the problem is that the peaks in $\gamma_{str}$ are quite narrow (which is a sign that the method is working well), so a small error in the location of the critical point results in a large error in $\gamma_{str}$. This problem should be less acute for larger $q$, where the peaks are broader. In any case, the method gives quite convincing results for the location of the critical point, which is what we want from it.

| $q$ | $U_3$ (log) | | | $U_4$ (log) | | | $U_3$ (no log) | | $U_4$ (no log) | |
|---|---|---|---|---|---|---|---|---|---|---|
| | $a_c$ | $\gamma_{str}$ | $\lambda$ | $a_c$ | $\gamma_{str}$ | $\lambda$ | $a_c$ | $\gamma_{str}$ | $a_c$ | $\gamma_{str}$ |
| 2 | | | | | | | 0.131 | −0.347 | 0.159 | −0.304 |
| 3 | | | | | | | 0.206 | −0.267 | 0.251 | −0.299 |
| 4 | 0.290 | 0.00329 | −0.710 | 0.352 | 0.0384 | −0.780 | 0.260 | −0.206 | 0.319 | −0.248 |
| 5 | 0.337 | 0.0359 | −0.711 | 0.408 | 0.0219 | −0.724 | 0.301 | −0.158 | 0.370 | −0.204 |
| 6 | 0.374 | 0.0741 | −0.726 | 0.454 | 0.0488 | −0.725 | 0.334 | −0.121 | 0.413 | −0.168 |
| 10 | 0.476 | 0.169 | −0.753 | 0.585 | 0.104 | −0.706 | 0.423 | −0.0231 | 0.530 | −0.0679 |
| 25 | 0.627 | 0.200 | −0.644 | 0.796 | −0.0278 | −0.421 | 0.553 | 0.107 | 0.714 | 0.0934 |
| 50 | 0.700 | 0.102 | −0.427 | 0.916 | −0.304 | −0.0345 | 0.614 | 0.158 | 0.812 | 0.196 |
| 100 | 0.737 | −0.0294 | −0.194 | 1.018 | −0.736 | 0.493 | 0.644 | 0.174 | 0.870 | 0.268 |
| 1000 | | | | | | | 0.665 | 0.166 | 0.925 | 0.360 |
| $\infty$ | | | | | | | 0.666 | 0.162 | 0.930 | 0.372 |

Table 3: Critical points and exponents calculated by the ratio method

Although it is not known exactly, one suspects that for $q = 4$ ($c = 1$) and higher, there are logarithmic corrections to scaling: that is, $\lambda \neq 0$ in eq. (22). Therefore I do the fit (32) with the logarithm. We have $\alpha = 0$ at $q = 4$ by KPZ, and $\alpha \approx 0$ at $q = 10$ by Monte Carlo; but we do not know whether $\alpha$ is also modified by logarithms or not. If it is, then following the discussion in section 3.1 one should use the peak in $U_4$ to locate the phase transition; but if $\alpha = 0$ is not modified by logarithms, one should use the $U_3$ peak instead. Consider the $q \geq 4$ region of Fig. 5. A striking result is the difference between the logarithmic and the non-logarithmic fits. At $q = 4$, both non-logarithmic fits give $\gamma_{str} \approx -0.2$, in agreement with Monte Carlo measurements, and in disagreement with KPZ. Just by including a logarithmic term one obtains radically better agreement with $\gamma_{str} = 0$. Since the $U_3$ $\gamma_{str}$ is an order of magnitude closer to zero than the $U_4$ $\gamma_{str}$, we have evidence that at $q = 4$, $\alpha = 0$ without logarithmic corrections. Thus for the $c = 1, q = 4$ model I conclude that $\gamma_{str} \approx 0$ in agreement with KPZ, but with logarithmic corrections: $\lambda \approx -0.7$. This should be compared to a Monte Carlo calculation for the same model which gives $\lambda \approx -1.5$, as well as another $c = 1$ model, the continuous $d = 1$ model, which has $\lambda = -1$.

For $q > 4$ the logarithmic corrections are likely to persist. Up until $q \approx 10$, all four estimates for $\gamma_{str}$ grow. The $U_3$ and $U_4$ non-logarithmic estimates follow each other closely, and cross zero around $q = 10 - 15$. The logarithmic approximations also increase, the $U_3$ faster than $U_4$. However, by $q \approx 15$ (for $U_4$) and $q \approx 25$ (for $U_3$) the $\gamma_{str}$ turns around, eventually reaching zero and becoming negative. The logarithmic exponent $\lambda$ stays negative and rather flat in the range $q \approx 4 - 10$, and then goes to zero, around where $\gamma_{str}$ begins to decrease. For the $U_3$ case, $\lambda$ seems to go to zero, while for the $U_4$ case it becomes positive. As I have already mentioned, logarithmic fits are dangerous when there is no logarithmic scaling, or when the logarithms are weak. This is precisely what happens here: when the logarithms start to weaken, the entire approximation starts to get erratic. It seems that in the range $q = 4 - 15$ or 20 there is a rather flat logarithmic correction, with the exponent $\lambda \approx -0.7$ or $-0.8$; after which the logarithmic correction decreases (it is impossible to say at present whether this decrease is gradual or sudden), and one can no longer trust the logarithmic fits.

Around $q \approx 20$ for the $U_4$ case, and $q \approx 40$ for the $U_3$ case, the logarithmic estimates for $\gamma_{str}$ cross their non-logarithmic counterparts. Since all indications are that the logarithmic corrections fade out by that point (remember, at $q = \infty$ we have $\lambda = 0$), we should trust the non-logarithmic estimates above $q \approx 40$. At this point the non-logarithmic $U_3$ estimate diverges from the $U_4$ estimate. As $q \to \infty$, we have $\gamma_{str}(U_3) \to 0.16$, while $\gamma_{str}(U_4) \to 0.37$. Only the latter fits well with the exact solution of the $q = \infty$ model [8], where $\alpha = -1$ (without logarithms: therefore the peak in $U_4$ should signal phase transition) and $\gamma_{str} = 1/3$ at the phase transition. All of this indicates that above $q \approx 40$ the non-logarithmic $U_4$ estimate for the critical point is the one to use.

It might seem worrisome that $\gamma_{str}$ overshoots its $q = \infty$ value $1/3$, settling down to the somewhat higher value 0.37. Similaly, $a_c \to 0.93$ (in place of the exact $a_c = 1$), and $g_c \to 0.0582$ (in place of the exact $1/\sqrt{288} \approx 0.05893$). This is, however, a finite-order artefact of the LTE. This can be checked by expanding the exact $q = \infty$ solution in LTE. What happens is that the approximants are slightly oscillatory (with period two: a common phenomenon, usually due to nearby unphysical singularities, and which is often cured by an Euler transform in the complex $a$-plane; but the cure can sometimes be worse than the disease), and don't completely settle down until $n > 10$. An *ad hoc* solution is to take the alternating approximants $a(8)$ and $a(10)$ and extrapolate them linearly in $1/n$ to $n \to \infty$, and thereby obtain a more refined estimate for $U$ and its derivatives. This gives a much better location for the critical point: $a_c = 1.011$, $g_c = 0.05898$.

If the magnetization phase transition were first-order (as it is on a flat surface for $q > 4$), the exponent $\gamma_{str}$ at the muticritical point should be exactly $-\frac{1}{2}$, the pure gravity value: only critical fluctuations should modify $\gamma_{str}$. Except possibly for the strange region around $q = 40$, $\gamma_{str}$ gives no hint of descending back down to $-\frac{1}{2}$, for any value of $q$. One therefore concludes that coupling a Potts model to a random surface softens the phase transition, making it continuous for most—if not all—values of $q$.[9]

## 4 Prospects

In this paper I have shown how to develop the LTE for any multi-matrix model on any target space. Computationally, the task is to calculate homomorphism coloring factors onto the desired target space. For the case of the Potts model, these homomorphism coloring factors are just the ordinary chromatic polynomials which are very easy to calculate. So for this model I have worked out the LTE out to 10th order. This takes about two hours of computer time on a workstation, and gives quite accurate results for not-so-branched ($q \leq 4$, $c \leq 1$) surfaces, as well as for very branched (large $q$, $c$) ones; and a plausible account of the mysterious region in between.

It is important to find ways to carry the LTE to higher order than 10th (the fact that it takes about 2 hours of computer time to go this far is promising in this respect). This is not just busywork. In ordinary spin models, especially with logarithmic corrections, one starts to get very accurate estimates for the exponents in order 15 or 20, typically. A similar effect can be seen in the $q = \infty$ model, which, since it can be solved exactly, can be expanded to arbitrary order in the LTE: the series is rather chaotic until order 6 or 7, then starts to behave well, but does not become completely smooth until order 15 or so. Thus it could be very profitable to push the LTE a few orders further. One way to do it would be to count only 1- or 2-particle irreducible skeletons; the full series can then be reconstructed.

[9] A possibly interesting comparison can be made to the Potts spin glass, which, for sufficiently large bond disorder, also has continuous transitions for all $q$ [28].

For example, of the 17,576 skeletons in 10th order, only about 200 are 2PI. Another improvement would be to calculate moments of the magnetization (rather than just closed surfaces or punctured ones with no spins on the punctures). One could then use the powerful Binder's cumulant technique to calculate the location of the critical point, rather than the peaks of the specific heat and its derivatives.

Having shown that the LTE works well for the Potts model on a random surface, nothing stands in the way of extending the calculations to other, more interesting, models. Well, almost nothing: one first has to devise a way of computing homomorphism coloring factors onto general target space graphs. But this is only a computational, rather than a conceptual, problem. Examples of models to study are $d$-dimensional lattices and multiple Ising models, which can easily implement $c > 1$ (for some recent work, see [24, 25, 23, 26, 27]. At $c = \infty$ these models are identical to the $q = \infty$ model [4], but the region of intermediate central charge is still poorly understood. The LTE seems to be a promising tool for its investigation.

I am grateful to LPTHE, Jussieu for its *accueil chalereux* and to Marco Picco for useful conversations of all kinds. The graphs used in this work were kindly provided by Brendan McKay.